\documentstyle[12pt]{article}
\newcommand{\beq}{\begin{equation}}
\newcommand{\eeq}{\end{equation}}
\newcommand{\bea}{\begin{eqnarray}}
\newcommand{\eea}{\end{eqnarray}}

\newcommand{\gsim}{\lower.7ex\hbox{$
\;\stackrel{\textstyle>}{\sim}\;$}}
\newcommand{\lsim}{\lower.7ex\hbox{$
\;\stackrel{\textstyle<}{\sim}\;$}}

\def\lsim{\mathrel{\rlap{\lower3pt\hbox{\hskip0pt$\sim$}}
    \raise1pt\hbox{$<$}}}         
\def\gsim{\mathrel{\rlap{\lower4pt\hbox{\hskip1pt$\sim$}}
    \raise1pt\hbox{$>$}}}         

\renewcommand{\Im}{{\rm Im}\,}

\newcommand{\bibit}[1]{\bibitem{#1}}

\newcommand{\La}{\overline{\Lambda}}
\newcommand{\Lam}{\Lambda_{\rm QCD}}

\newcommand{\as}{\alpha_s}
\newcommand{\GeV}{\,\mbox{GeV}}

\newcommand{\matel}[3]{\langle #1|#2|#3\rangle}

\begin{document}

\begin{titlepage}
\renewcommand{\thefootnote}{\fnsymbol{footnote}}

\begin{flushright}
UND-HEP-00-BIG\hspace*{.03em}09\\
hep-ph/0011124\\
\end{flushright}
\vspace*{1.3cm}

\begin{center} \Large
{\bf New Exact Heavy Quark Sum Rules}
\end{center}
\vspace*{1.5cm}
\begin{center} {\Large
Nikolai Uraltsev} \\
\vspace{1.2cm}
{\normalsize
{\it Dept.\ of Physics, Univ.\ of Notre Dame du Lac, Notre Dame, IN 46556,
U.S.A.$^*$}\\
{\small \rm and} \\
{\it INFN, Sezione di Milano, Milan, Italy$^{\:*}$
}\\
}

\normalsize
\vspace*{1.7cm}

\vspace*{1.4cm}

{\large{\bf Abstract}}\\
\end{center}
\vspace*{.2cm}
Considering nonforward scattering amplitude off the heavy quark in the
Small Velocity limit two exact superconvergent sum rules are derived. 
The first sum rule leads to the lower bound $\varrho^2\!>\! 3/4$ for the
slope of the Isgur-Wise function. It also provides the rationale for the
fact that the vector heavy flavor mesons are heavier than the
pseudoscalar ones. A spin-nonsinglet analogue $\overline\Sigma$ of 
$\La\!=\!M_B\!-\!m_b$ is introduced.

\vfill

~\hspace*{-7mm}\hrulefill \hspace*{3cm} \\
\footnotesize{\noindent $^*$On leave of absence from 
St.\,Petersburg Nuclear Physics Institute, Gatchina, St.\,Petersburg 
188300, Russia}

\noindent
\end{titlepage}

\newpage

Heavy quark symmetry and the heavy quark expansion have played an
important role in understanding weak decays of heavy flavors. Recent
years witnessed significant success in quantifying strong
nonperturbative dynamics in a number of practically important problems
via application of Wilson Operator Product Expansion (OPE). It is fair
to note, however, that with the analytic solution of QCD still missing,
the effect of nonperturbative low-scale domain is more parametrized
than computed from the first principles. Therefore, any model-independent
constraint on the nonperturbative parameters playing the role in heavy
quark decays, is an asset. A number of such relations come from the
heavy quark sum rules, in particular for transitions between two
sufficiently heavy quarks with no change of velocity (zero-recoil sum
rules), or where velocity changes by a small amount (small velocity, or
SV sum rules). The unified derivation of the sum rules in the
field-theoretic OPE is described in detail in the dedicated papers
\cite{optical}, with their quantum mechanical interpretation 
elucidated. A more pedagogical derivation can be found in recent reviews
\cite{rev,ioffe}.\footnote{An interesting introduction to heavy quarks
in QCD can be found in lectures \cite{shifman}, with the references
therein to the more conventional reviews.}

In this letter the standard OPE approach is extended to the nonforward
SV scattering amplitude of weak currents off the heavy quark. This leads
to two new exact heavy quark sum rules related to the spin of
the light constituents of the heavy flavor hadron.

We consider the SV scattering amplitude 
\beq
T (q_0; \vec{v},\vec{u}) =
\frac{1}{2M_{H_Q}}
\matel{H_Q(\vec{u})}{\int \; {\rm d}^3x\,{\rm d}x_0\;
{\rm e}\,^{i\vec{q}\vec x -iq_0x_0}\:
iT\{J_0^\dagger(0),\, J_0(x)\}  }{H_Q(0)},
\label{7}
\eeq
where $J_\mu \!=\!\bar{Q}\gamma_\mu Q$ and $\vec{q} \!\equiv\! m_Q
\vec{v}$; likewise $m_Q\vec{u}$ is the momentum of the final heavy
flavor hadron. We will also denote $\vec{v}'\!=\!\vec{v}\!-\!\vec{u}$.
Note that $J_\mu$ is the nonrelativistic current of heavy quarks, and
the field $Q(x)$ entering it includes only the operator of annihilation
of the heavy quark contained in $H_Q$, but not the creation of $\tilde
Q$ which is present in $ Q(x)$ from $J_0^\dagger$. 
Therefore, $J_\mu^\dagger$ is
different from $J_\mu$ and only the product in the order
$J_0^\dagger(0) J_0(x)$ contributes (we adopt the convention where $H_Q$
contains heavy quark and not the antiquark). This can be visualized
considering the nondiagonal $b\!\to\! c$ transitions with $J_\mu
\!=\!\bar{c}\gamma_\mu b$, $J_0^\dagger\!=\!\bar{b}\gamma_\mu c$,
however we assume that both $m_b$, $m_c \to \infty$ and, for simplicity,
put $m_b\!=\!m_c$. 

In the SV limit we retain only terms through second order in $\vec v$
and $\vec u$ and take the energy variable $\epsilon$ according to 
\beq
\epsilon=q_0-\left(\sqrt{\vec{q}^{\:2}\!+\!m_Q^2}\!-\!m_Q \right)\simeq
q_0\!-\!\frac{m_Q\vec{v}^{\:2}}{2}\;;
\label{9}
\eeq
the elastic transitions for a free quark would then correspond to
$\epsilon\!=\!0$. The amplitude $T (\epsilon; \vec{v},\vec{v}\!-\!\vec{v}')$ 
can be
decomposed into symmetric and antisymmetric in $\vec v$, $\vec v'$ parts
$h_+(\epsilon)$ and $h_-(\epsilon)$; the latter is present if $H_Q$ has
nonzero spin of light degrees of freedom $j$ correlated with its total
spin, {\it viz.\ }$h_- \sim i\epsilon_{kln}v_k v'_l
\matel{H_Q}{j_n}{H_Q}$. 

At large (complex) $\epsilon \!\gg \! \Lam$ the amplitude (\ref{7}) can
be expanded in inverse powers of $\epsilon$. Simultaneously we assume
that $\epsilon \!\ll \! m_Q$ and can discard all terms suppressed by
powers of $m_Q$. In this limit the OPE simplifies and takes the form
\bea
\nonumber
-T (\epsilon; \vec{v},\vec{u}\,)
\!\!\!\!&=&\!\!\!
\frac{1}{\epsilon}_{\,} \;\frac{1}{2M_{H_Q}}
\matel{H_Q(\vec{u}\,)}{\,\bar{Q} (
1\!-\!\mbox{{\small $\frac{\vec
v^{\,2}}{4}$}}\!+\!\mbox{{\small $\frac{\vec{v}\vec{\gamma}}{2}$}}
\,) Q(0)\,}{H_Q(0)}\\
\nonumber
&+&\!\!\!\!
\frac{1}{\epsilon^2} \; \frac{1}{2M_{H_Q}}
\matel{H_Q(\vec{u}\,)}{\,\bar{Q} (i\vec{D}\vec{v}\,)
Q(0)}{H_Q(0)}\\
+ \;\sum_{k=0}^{\infty} \!\!\!\!\!&&\!\!\!\!\!\!\!
\frac{(\!\mbox{-}_{\!}1\!)^k\!}{\epsilon^{k\!+\!3}} \, \frac{1}{2M_{H_{Q\!}}}
\matel{H_Q(\vec{u}\,)}{\,\bar{Q} [(i\vec{D}\vec{v}\,)\pi_0^k
(i\vec{D}\vec{v}\,)
\!-\!
\pi_0^{k+\!1}(i\vec{D}\vec{v}\,)]
Q(0)}{H_Q(0)}.
\label{13}
\eea
Here $\pi_0$ is nonrelativistic energy, $\pi_0\!=\!iD_0\!-\!m_Q$.

On the other hand, the dispersion relation equates the coefficients in
the $1/\epsilon$ series to the moments of the absorptive part of
$T(\epsilon; \vec{v},\vec{u}\,)$ (nonforward structure functions of the
hadron $H_Q$):
\beq
-T (\epsilon; \vec{v},\vec{u}\,)=
\sum_{k=0}^{\infty} \frac{1}{\epsilon^{k\!+\!1}} \:
\frac{1}{2\pi} \int {\rm d}\omega \:\omega^k \;\Im 
T(\omega;\vec{v},\vec{u}\,)\;;
\label{15}
\eeq
the latter is given by the product of the SV transition amplitudes into
the ground and excited stated. $T$ is analytic at negative $\epsilon$
and has a cut at positive $\epsilon$. The elastic transitions lead to
the pole located at $\epsilon\!\simeq\! -\frac{\La\vec{v}^{\:2}}{2}$.
Equating $1/\epsilon^k$ terms in Eqs.~(\ref{13}) and (\ref{15}) we get
the sum rules. We illustrate them on the example of B mesons 
($j\!=\!\frac{1}{2}$) taking $B^*$ as $H_Q$. With $j\!=\!\frac{1}{2}$ the only
nontrivial symmetric structure in $h_+$ is proportional to 
$(\vec{v}\vec{v}_{\!}\,') \!\cdot\! \langle
H_Q'|H_Q\rangle$. Therefore, it is convenient to introduce 
\vspace*{-1mm}
$$
-T (\epsilon; \vec{v},\vec{v}\!-\!\vec{v}_{\!}\,') \; = \;
\left[
\frac{1\!-\!a(\vec{v}^{\,2}\!+\!\vec{v}_{\!}\,'^2)\!-\!
b(\vec{v}\!-\!\vec{v}_{\!}\,')^2}{\epsilon}
\!-\! \frac{c\,\vec{v}^{\:2}}{\epsilon^2} 
+
(\vec{v}_{\,}\vec{v}_{\!}\,') \,h_+(\epsilon) \right]
\mbox{\large $\frac{\langle H_Q'|H_Q\rangle}{2M_{H_Q}}$}
\qquad\qquad
$$
\vspace*{-2mm}
\beq
\qquad\qquad\qquad\qquad -\; 
i\epsilon_{jkl}\,v_j \,v'_k  \: h_-(\epsilon) \:
\mbox{\large $ \frac{\matel{H_Q'}{J_l}{H_Q}}{2M_{H_Q}}$} 
\;+ \;{\cal O}(\!\vec{v}^{\,3})
\label{17}
\eeq
\vspace*{-1mm}
and
\vspace*{-.5mm}
\beq
W_+(\epsilon)= \frac{1}{2\pi} \Im h_+(\epsilon)\;,\qquad 
W_-(\epsilon)= \frac{1}{2\pi} \Im h_-(\epsilon)
\label{19}
\eeq
with 
$\vec J$ denoting the 
angular momentum of $H_Q$; 
constants $a,\,b$ and $c$ are associated with the
elastic transition.

The SV transitions can proceed  to $j\!=\!\frac{1}{2}$ (scalar $S$ and
axial $A$)  and to $j\!=\!\frac{3}{2}$ (axial $D_1$ and tensor $D_2$)
``$P$-wave'' states. Following the notations of Ref.~\cite{isgw}, the 
corresponding nonrelativistic amplitudes are
given by 
\bea
\nonumber
\matel{S(\vec v_2)}{J_0}{B^*(\varepsilon,\vec v_1)}
\!\!\!\!&=&_{_{_{\!\,}}}
\tau_{1/2} \left(\vec{\varepsilon}_{\,} 
(\!\vec v_2\!-\!\vec v_1\!)\right)\;,\\
\matel{A(e,\vec v_2)}{J_0}{B^*(\varepsilon,\vec v_1)}
\!\!\!\!&=&\!\!\!\!
-\tau_{1/2}\, i\epsilon_{jkl}\, e^*_j \,\varepsilon_k \,(\! v_2\!-\!v_1\!)_l
\label{21}
\eea
\vspace*{-1mm}
and
\vspace*{-1mm}
\bea
\nonumber
\matel{D_1(e,\vec v_2)}{J_0}{B^*(\varepsilon,\vec v_1)}
\!\!\!\!&=&\!\!\!\!
-\frac{1}{\sqrt{2}}\,\tau_{3/2}\, 
i\epsilon_{jkl}\, e^*_j \,\varepsilon_k \, (\!v_2\!-\! v_1\!)_l\;,\\
\matel{D_2(e,\vec v_2)}{J_0}{B^*(\varepsilon,\vec v_1)}
\!\!\!\!&=&\:
\sqrt{3}_{\;}\tau_{3/2} \,e^*_{jk} \,\varepsilon_j(v_2\!-\!v_1)_k\;.
\label{23}
\eea
Here $\varepsilon$, $e$ are nonrelativistic 3-vectors of polarization.
In the case of $D_2$, however $e$ is the symmetric rank-$2$ tensor, with
the sum over polarizations 
\beq
\sum_\lambda e^{(\lambda)}_{ij} e^{(\lambda)}_{kl} = 
-\frac{1}{3} \delta_{ij} \delta_{kl} + 
\frac{1}{2} (\delta_{ik}\delta_{jl} + \delta_{il}\delta_{jk})\;.
\label{25}
\eeq
The elastic transition to this order can proceed only to $B^*$, and the
amplitude is given by
\beq
\matel{B^*(\vec\varepsilon_2,\vec v_2)}{J_0}{B^*(\vec\varepsilon_1,\vec v_1)}
= 2M_{B^*} \cdot \xi \!\left(\!(\vec{v}_2\!-\!\vec{v}_1\!)^{2\!}\right) 
\left(1\!+\!\mbox{\small 
$\frac{\vec v_1^{\,2} \!+  \vec v_2^{\,2}}{4}$}\right)
(\vec\varepsilon_2^{\,*} \vec\varepsilon_1) + {\cal O}(\vec{v}^{\:3})\;,
\label{27}
\eeq
where $\xi$ is the Isgur-Wise function (its slope
$\varrho^2\!=\!-2\xi'(0)$ will be used later) 
and $\vec\varepsilon$ are the rest-frame 
polarizations. The latter are
related to the polarization $4$-vectors $\epsilon$ via
$\epsilon_0\!=\!(\vec{v}\vec\varepsilon)$, $\vec\epsilon\!=\! 
\vec\varepsilon +
\frac{1}{2} \vec{v}(\vec{v}\vec\varepsilon)$, up to terms cubic in
velocity. 

The explicit computation yields the following structures for the 
contributions 
of the $\frac{1}{2}$ and $\frac{3}{2}$ multiplets, 
\bea
\nonumber
&&|\tau_{1/2}|^2 \left\{ \,\;\,_{_{\!\!}}
(\vec\varepsilon\,'^* \vec \varepsilon\,)(\vec v'\,\vec v)
- \left[
(\vec\varepsilon\,'^* \vec v\,)(\vec\varepsilon\,\vec v\,'\,)
\!-\!
(\vec\varepsilon\,'^*\vec v\,')(\vec\varepsilon \,\vec v\,)
\right]
\right\} ,\\
&& |\tau_{3/2}|^2 \left\{\!\,
2(\vec\varepsilon\,'^* \vec \varepsilon\,)(\vec v'\,\vec v)
+ \left[
(\vec\varepsilon\,'^* \vec v\,)(\vec\varepsilon\,\vec v\,'\,)
\!-\!
(\vec\varepsilon\,'^*\vec v\,')(\vec\varepsilon \,\vec v\,)
\right]
\right\},
\label{31}
\eea
respectively, where $\vec\varepsilon\,'$ is the polarization vector of
$B^*(\vec u\,)$. We also need the following matrix elements to evaluate
the OPE part:
\beq
\frac{1}{2M_{B^*}\!}\matel{H_Q(\varepsilon'\!\!,\vec{u}_{\,})}{\bar{Q} Q(0)}
{H_Q(\varepsilon,0)} =
\xi(\vec{u}^{\:2})(1\!+\!\mbox{$\frac{\vec{u}^{\,2\!}}{4}$}) 
\,(\vec\varepsilon\,'^* \vec \varepsilon\,)\;
,
\label{32}
\eeq
\beq
\frac{1}{2M_{B^*}\!}\matel{H_Q(\varepsilon'\!\!,\vec{u}_{\,})}{\bar{Q}
\gamma_i Q(0)}
{H_Q(\varepsilon,0)} =
 \mbox{$\frac{1}{2}$} u_i\:(\vec\varepsilon\,'^* \vec \varepsilon\,)
+ \mbox{$\frac{1}{2}$} \left[
(\vec\varepsilon\,'^* \vec u\,) \varepsilon_i - \varepsilon'^*_i(\vec
\varepsilon\,\vec u\,)
\right]
\label{33}
\eeq
and 
\beq
\frac{1}{\!2M_{B^*}\!\!}
\matel{H_Q(\varepsilon'\!\!,\vec{u}^{\,})}{\bar{Q} iD_{\!j} Q(0)}
{H_Q(\varepsilon,0)} =
- \frac{\La}{2}\, u_j \:(\vec\varepsilon\,'^* \vec \varepsilon\,)
+
\frac{\overline\Sigma}{2} \!
\left\{\!(\vec\varepsilon\,'^* \vec u^{\,})\varepsilon_j
\!-\!
\varepsilon'^*_j(\vec\varepsilon^{\,} \vec u^{\,})\!\right\}\:
 +
{\cal O}\!\left(\!\vec u^{\,2\!}\right)^{\!}
.
\label{35}
\eeq
The higher order in $\vec u,\:\vec v$ terms have been omitted from all
the expressions. The last matrix element introduces the new hadronic
parameter $\overline \Sigma$ which can be viewed as the spin-nonsinglet
analogue of $\La\!=\!\lim_{m_b\!\to\!\infty} M_B\!-\!m_b$. 
The part proportional to $u_j$ in Eq.~(\ref{35}) amounts to
$-\frac{\La}{2}$; this follows from the nonrelativistic equation of motion
$(v_\mu iD_\mu) Q(x)\!=\! m_Q Q(x)$ which holds in the static limit
$m_Q\!\to\!\infty$ for the heavy quark moving with velocity $v$:
$$
M_{H_Q} ( u\!-\!u^{(0)})_\mu u_\mu
\matel{H_Q(\vec{u}\,)}{\bar{Q} Q(0)}{H_Q(0)}=
\left.
-(u_\mu iD_\mu) \,\matel{H_Q(\vec{u}\,)}{\bar{Q}
Q(x)}{H_Q(0)}\right\vert_{x\!=\!0}\;=
$$
$$
\matel{H_Q(\vec{u}\,)}{m_Q \bar{Q}Q(0)- \bar{Q}(u_\mu iD_\mu) Q(0)}{H_Q(0)}
$$
($u^{(0)}=(1,\vec{0})$ is the restframe four-velocity) which to order 
$\vec{u}^{\,2}$ leads to 
\beq
\left(M_{H_Q}\!-\!m_Q\right) \frac{\vec{u}^{\,2}}{2} \:
\matel{H_Q}{\bar{Q} Q(0)}{H_Q} = - u_k
\matel{H_Q(\vec{u}\,)}{\bar{Q} iD_k Q(0)}{H_Q(0)}\;.
\label{39}
\eeq

Considering the forward scattering amplitude where $\vec{u}\!=\!0$ and
$\vec v\,'\!=\!-\vec v$ selects the symmetric structure function
$W_+(\epsilon)$ and yields the Bjorken \cite{BJSR} and Voloshin
\cite{volopt} sum rules for the
zeroth and first moments:
\bea
\varrho^2\!-\!\frac{1}{4} \!\!&=&
 2\sum_m\;\;|\tau_{3/2}^{(m)}|^2 \,\;\;+\;\;\;
\sum_n\;|\tau_{1/2}^{(n)}|^2 \, ,
\label{41}
\\
\frac{\La}{2}\; &=&
2\sum_m \epsilon_m|\tau_{3/2}^{(m)}|^2 \;+\;\;\;
\sum_n \epsilon_n|\tau_{1/2}^{(n)}|^2\, ,
\label{43}
\eea
with $\epsilon_k\!=\!M^{(k)}\!-\!M_B$ denoting the mass gap between the
$P$-wave and the ground state. Setting $\vec{v}\!=\!0$ or 
$\vec{_{\,}v}_{\!}\,'\!\!=\!0$ 
fixes
$a\!=\!\varrho^2/2$, 
$b\!=\!-\frac{1}{4}$
and $c\!=\!\frac{\La}{2}$ in Eq.~(\ref{17}), as expected.

The new sum rules emerge for the zeroth and first moments of the 
antisymmetric structure function $W_-$:
\bea
\frac{1}{4}\,\: &=&
 \sum_m\;|\tau_{3/2}^{(m)}|^2 \,\;\;\;-\:
 \sum_n\;|\tau_{1/2}^{(n)}|^2 \, ,
\label{45}
\\
\frac{\overline \Sigma}{2}\; &=&
\sum_m \epsilon_m|\tau_{3/2}^{(m)}|^2 \;-\;
\sum_n \epsilon_n|\tau_{1/2}^{(n)}|^2\, .
\label{47}
\eea
The higher moments yield the known sum rules \cite{rev} for $\mu_G^2$,
$\rho_{LS}^3$, \ldots which can be obtained, for instance, considering
the usual zero-recoil structure functions of spacelike components of vector
currents appearing at order $1/m_Q^2$ \cite{optical}. 

It is often convenient to work in the static limit $m_Q\!\to\!\infty$
assuming that $Q$ are spinless; $B$ and $B^*$ in this case are different
components of a single spin-$\frac{1}{2}$ particle, $\Psi_0$. The derivation
of the sum rules in this case proceeds similarly; the antisymmetric
structure would be absent from the OPE, but the elastic transition yield
it with the opposite sign for the zeroth moment. The $D\!=\!4$ matrix
element in this case takes the form
\beq
\matel{\Psi_0(\vec{u})}{\bar{Q} iD_j Q(0)}{\Psi_0(0)} =
-\frac{\La}{2}\, u_j\, \Psi_0^\dagger \Psi_0 - i
\frac{\overline{\Sigma}}{2} \,\epsilon_{jkl} \,u_k \,
\Psi_0^\dagger \sigma_{l} \Psi_0+
{\cal O}\left(\vec u^{\,2}\right)
\;.
\label{51}
\eeq

The exact magnitude of the nonperturbative hadronic parameter 
$\overline\Sigma$ is not
known at the moment, but can be estimated by means of QCD sum rules; or
it can be directly measured on the lattices. Comparing the sum rules
(\ref{45}), (\ref{47}) with the sum rule for the chromomagnetic
expectation value $\mu_G^2$ \cite{rev}
\beq
\frac{\mu_G^2}{6} =
\sum_m \epsilon_m^2|\tau_{3/2}^{(m)}|^2 \;-\;
\sum_n \epsilon_n^2|\tau_{1/2}^{(n)}|^2 \, ,
\label{53}
\eeq
$$
\mu_G^2=\frac{1}{2M_{H_Q}} \matel{B}{\bar{b}
\mbox{$\frac{i}{2}$}\sigma_{\mu\nu}G^{\mu\nu} b(0)}{B}\simeq
\frac{3}{4}\,(M^2_{B^*} \!-\! M^2_B)\approx 0.4\GeV^2\;,
$$
we expect $\overline\Sigma$ to be about $0.25\GeV$.

The first sum rule (\ref{45}) which is independent of the strong
dynamics at first may look surprising. In the quark models the
$\frac{1}{2}$ and $\frac{3}{2}$ states are differentiated only by
spin-orbital interaction. The latter naively can be taken arbitrarily
small if the light quark in the meson is nonrelativistic. To
resolve this apparent paradox we note that in the
nonrelativistic case $\tau^2$ are large scaling like inverse square of
the typical velocity of the light
quark, $\tau^2\sim 1/\vec v_{\rm sp}^{\,2}$, and the relativistic
spin-orbital
effects must appear at the relative level $\sim \vec v_{\rm sp}^{\,2}$
because spin ceases to commute with momentum to this accuracy. These
relativistic corrections lead to the terms of order $1$ in the first 
sum rules Eqs.\,(\ref{41}), (\ref{45}). 
The constant $\frac{1}{4}$ comes in the latter case from the
$1/m^2$ $LS$-term; this is easy to show using the commutation relations
between the momentum, coordinate and the nonrelativistic Hamiltonian.

The heavy quark sum rules lead to a number of exact inequalities in the
static limit \cite{rev,ioffe}. The most familiar one is the Bjorken
bound $\varrho^2 \!>\! 1/4$. The sum rule (\ref{45}) gives us the stronger
dynamical bound $\varrho^2 \!>\! 3/4$. Comparison to the QCD sum rule
evaluation $\varrho^2 \!=\! 0.7\!\pm\! 0.1$ \cite{bloks} suggests that
this bound can be nearly saturated.
We also have the bound $\La\!>\! 2\overline\Sigma$. 

The sum rule (\ref{45}) provides the rationale for the experimental fact
that vector mesons $B^*$, $D^*$ are heavier than their
hyperfine pseudoscalar partners $B$, $D$. Indeed, if the sum rule for
$\mu_G^2$ is dominated by
the low-lying states then $\mu_G^2$ must be of the same sign as the
constant in Eq.~(\ref{45}), which dictates the negative energy of the
heavy quark spin interaction in $B$ and positive in $B^*$.

Let us note that the full matrix element in Eqs.~(\ref{35}), (\ref{51})
is not well defined in the limit $m_Q\!\to\!\infty$ due to ultraviolet
divergences in the static theory, and therefore the spin-independent
part proportional to $\La$ depends on regularization. Nevertheless, the
antisymmetric part proportional to $\overline{\Sigma}$ is well defined
and finite. The new sum rules (\ref{45}) and (\ref{47}) are convergent
and not renormalized by perturbative corrections; this distinguishes them
from all other heavy quark sum rules. 

At large $\epsilon\!\gg \! \Lam$ the sums over the excited states are
dual to the quark-gluon contribution computed in perturbation theory
\cite{varenna,ioffe}:
\bea
2\sum_{m} \epsilon_m^2|\tau_{3/2}^{(m)}|^2 +
\sum_{n} \epsilon_n^2 |\tau_{1/2}^{(m)}|^2 \!\!& \to &
\frac{8}{9}\frac{\as(\epsilon)}{\pi}\: \epsilon\,  {\rm d}\epsilon\,
\;,
\label{71}
\\
\sum_{m}
\epsilon_m^2|\tau_{3/2}^{(m)}|^2 -
\sum_{n}
\epsilon_n^2|\tau_{1/2}^{(n)}|^2  \!&\to & \!
-\frac{3\as(\epsilon)}{2\pi}\, \frac{{\rm d}\epsilon\,}{\epsilon}
\left\{
\sum_{\epsilon_m <\epsilon} \epsilon_m^2|\tau_{3/2}^{(m)}|^2 \!-\!
\sum_{\epsilon_n <\epsilon} \epsilon_n^2|\tau_{1/2}^{(n)}|^2
\!\right\}, \;\;\:\qquad
\label{73}
\eea
where the embraced expression simply amounts 
to $\frac{1}{6} \mu_G^2(\epsilon)$
if the normalization point is implemented as the cutoff in the energy
$\epsilon$. Equation (\ref{71}) can be immediately extended to higher
orders in $\alpha_s$, this amounts to using the so-called dipole
coupling $\alpha_s^{(d)}(\epsilon)$ introduced in Ref.~\cite{dipole}:
\beq
\alpha_s^{(d)}(\epsilon) =
\bar\alpha_s\left({\rm e}^{-5/3+\ln{2}} \epsilon\right)
- \mbox{\large $\left(\frac{\pi^2}{2}\!-\!\frac{13}{4} \right)
\frac{\as^2}{\pi}$} \,+\,{\cal O}(\as^3)\;.
\label{75}
\eeq
Using Eq.~(\ref{73}) we can estimate the contribution of the high-energy
states in the sum rules (\ref{45}) and (\ref{47}):
\bea
\sum_{\epsilon_m<\mu} \;\;\;|\tau_{3/2}^{(m)}|^2 -
\sum_{\epsilon_n<\mu} \;\;\,|\tau_{1/2}^{(m)}|^2 \!\! &\simeq & \!\!
_{\,}\frac{1}{4}_{\,}+ \frac{\as(\mu)}{8\pi}\: \frac{\mu_G^2(\mu)}{\mu^2}\;,
\label{77}
\\
\sum_{\epsilon_m<\mu} \!\epsilon_m |\tau_{3/2}^{(m)}|^2 -
\sum_{\epsilon_n<\mu} \!\epsilon_n |\tau_{1/2}^{(m)}|^2 \!\! &\simeq & \!\!
\frac{\overline\Sigma}{2}+ \frac{\as(\mu)}{4\pi}\:
\frac{\mu_G^2(\mu)}{\mu}\;;
\;\;\;\;
\label{79}
\eea
they are power suppressed and presumably small in the perturbative
domain.

The transition amplitudes with change of the heavy quark velocity
acquire ultraviolet divergences when $m_Q\!\to\!\infty$, being
suppressed by the nonrelativistic analogue of the Sudakov formfactor
\vspace*{-3mm}
\beq
S \;\sim\;  \mbox{e\Large$^{-\frac{4}{9\pi}(\Delta \vec v\,)^2
\int \frac{{\rm d} \omega}{\omega}\,\alpha_s^{(d)}(\omega)
}$}\;,
\label{83}
\eeq
\vspace*{-6mm}\\
where the upper limit of integration is set by $m_Q$. 
The nonforward scattering amplitude has the similar universal
suppression. This does not affect our consideration since these effects
are given only by symmetric combinations of $\vec{v}$ and $\vec{v}\,'$.

Even the convergent sum rules can, in principle, get a finite perturbative
renormalization as it happens, for example, with the Bjorken or 
Ellis-Jaffe sum rules in DIS where it comes from gluon momenta scaling
like $\sqrt{Q^2}$. Since we defined the SV transition amplitudes
($\tau$'s) via the flavor-diagonal vector current, such perturbative 
renormalization is absent from the sum rules. Alternatively, if the 
axial and/or flavor-nondiagonal currents are used, there will be the 
overall factor $\eta^2$ with $\eta$ the short-distance
renormalization of the corresponding zero recoil weak current which would 
enter the right hand sides of Eqs.~(\ref{21}) and (\ref{23}).

The heavy quark sum rules can also be considered for other heavy flavor
hadrons. In $\Lambda_b$ the spin of light degrees of freedom vanishes, 
and no nontrivial relations are obtained with the scattering amplitudes
at $\vec{u}\!\ne\!0$. They are informative for the $\Sigma_b$-type
states with $j\!=\!1$ and can be derived directly applying the quoted
equations to the corresponding states. 
Although for 
spin-$1$ light cloud an additional, spin-dependent amplitude is present
in the elastic transitions at order $\vec{v}^{\:2}$, it does not yield
the antisymmetric in $\vec{v}, \vec{v}_{\!}\,'$ structure. Since the
nonforward amplitude has three tensor structures bilinear in 
$\vec{v}, \vec{v}_{\!}\,'$, there are three sum rules for the
contributions of the three types of $P$-wave states proportional to
$|\tau_0|^2$, $|\tau_1|^2$ and $|\tau_3|^2$, respectively, which
constrain the slope of the Isgur-Wise function and the second formfactor
at zero recoil.

The sum rule (\ref{45}) can be applied in atomic physics; there it is
a sum rule for spin-orbital interaction in dipole transitions. There is
a difference between the transitions in atom and $B$ meson, though: in
the latter case we are studying the amplitudes mediated by the heavy
quark currents, whereas in atoms photons are emitted through their interaction
with electrons. The two amplitudes are directly related only in the
nonrelativistic approximation. In particular, additional relativistic
effects emerge due to explicit corrections in the electromagnetic current
of electrons.

The various sum rules we discuss for heavy flavor
transitions are the operator relations for the equal time commutators of
infinitely heavy quark currents (the lowest sum rules), or their time
derivatives, for higher sum rules. 
The OPE allows straightforward calculation of these commutators including
possible Schwinger terms.
\vspace*{.5mm}

{\it To summarize}, applying the OPE to the nonforward heavy quark
scattering amplitude two new superconvergent sum rules are derived
intrinsically connected to the spin of light cloud in the heavy flavor
hadron. The first sum rule leads to the nontrivial bound $\varrho^2\!>\!
\frac{3}{4}$ for the slope of the IW function, and suggests why $B^*$ is
heavier than $B$. The second sum rule
bounds the difference $M_B\!-\!m_b$ from below. These sum rules are exact in
the heavy quark limit and can help to constrain a number of
nonperturbative parameters in heavy flavor hadrons.
\vspace*{.2cm}\\
{\bf Acknowledgments:} ~\,Useful discussion with I.~Bigi  
and, in particular, M.~Eides, M.~Shifman and A.~Vainshtein are gratefully 
acknowledged.
This work was supported in part by NSF under grant number PHY96-05080 and by
RFFI under grant No.\ 99-02-18355.

\end{document}